\documentclass[10pt,aps,prb,amsmath,amssymb,floatfix,twocolumn,superscriptaddress]{revtex4-2}

\usepackage[dvipsnames]{xcolor}
\definecolor{webblue}{HTML}{2D3092}
\usepackage[
colorlinks,
linkcolor=webblue,
citecolor=webblue,
urlcolor=webblue
]{hyperref}
\usepackage[capitalize]{cleveref}
\usepackage{mathtools}
\usepackage{bm}
\usepackage{graphicx}
\usepackage{comment}
\usepackage{braket}
\usepackage{chemformula}

\newcommand{\supplement}[1]{%
  \clearpage%
  \title{#1}%
  \maketitle%
  \setcounter{section}{0}
  \setcounter{equation}{0}%
  \setcounter{figure}{0}%
  \setcounter{table}{0}%
  \setcounter{page}{1}%
  \makeatletter%
  \renewcommand{\thesection}{S\arabic{section}}%
  \renewcommand{\thesubsection}{\Alph{subsection}}%
  \renewcommand{\theequation}{S\arabic{equation}}%
  \renewcommand{\thefigure}{S\arabic{figure}}%
  \renewcommand{\thetable}{S\Roman{table}}%
  \renewcommand{\thepage}{S\arabic{page}}%
  \numberwithin{figure}{section}%
  \numberwithin{table}{section}%
  \numberwithin{equation}{section}%
  \makeatother%
  \onecolumngrid%
}

\makeatletter
\def\maketitle{
\@author@finish
\title@column\titleblock@produce
\suppressfloats[t]}
\makeatother

\newcommand{\prlparagraph}[1]{\textit{#1}---}
\newcommand{\nodag}{{\vphantom\dagger}}

\renewcommand{\Im}{\mathrm{Im}}

\newcommand{\Tr}{\text{Tr}}

\newcommand{\bS}{\mathbf{S}}
\newcommand{\bk}{\mathbf{k}}
\newcommand{\br}{\mathbf{r}}
\newcommand{\bq}{\mathbf{q}}

\begin{document}

\title{Altermagnons at the metal-insulator transition}

\author{Jonas Issing}
\email{jonas.issing@uni-wuerzburg.de}
\affiliation{Institut f\"ur Theoretische Physik und Astrophysik and W\"urzburg-Dresden Cluster of Excellence ctd.qmat, Julius-Maximilians-Universit\"at W\"urzburg, Am Hubland, Campus S\"ud, W\"urzburg 97074, Germany}

\author{Matteo Dürrnagel}
\author{Sarbajit Mazumdar}
\author{Alena Lorenz}
\author{Niklas Witt}
\author{Giorgio Sangiovanni}
\affiliation{Institut f\"ur Theoretische Physik und Astrophysik and W\"urzburg-Dresden Cluster of Excellence ctd.qmat, Julius-Maximilians-Universit\"at W\"urzburg, Am Hubland, Campus S\"ud, W\"urzburg 97074, Germany}

\author{Michael Klett}
\author{Lennart Klebl}
\author{Ronny Thomale}
\author{Jannis Seufert}
\email{jannis.seufert@uni-wuerzburg.de}
\affiliation{Institut f\"ur Theoretische Physik und Astrophysik and W\"urzburg-Dresden Cluster of Excellence ctd.qmat, Julius-Maximilians-Universit\"at W\"urzburg, Am Hubland, Campus S\"ud, W\"urzburg 97074, Germany}

\date{\today}
\begin{abstract}
 By means of slave-boson theory for the Hubbard model on the checkerboard lattice, we calculate dynamical altermagnetic spin susceptibilities from the metallic to the Mott-insulating regime. We track magnon dispersion and lifetime renormalization, allowing us to uncover a crossover from a chirality-selective dissipation of magnon modes to coherent yet strongly deformed chiral magnon branches across the metal insulator transition. Our formalism lends itself to a quantitative description of collective spin dynamics in correlated altermagnets. 
\end{abstract}
\maketitle

\prlparagraph{Introduction}%
In the past decades, spintronics has evolved into a vibrant area of condensed matter physics that concerns itself with the control, processing, and manipulation of spin signals in a solid-state setting~\cite{RevModPhys.76.323}. Owing to the variety of quasiparticles that carry spin, spintronics naturally divides into electron-based spintronics, where electrons are the source of the spin profile, and magnonics, where a similar role is picked up by spin wave excitations~\cite{pirro}. Magnonics attracts significant attention by promising novel devices for signal processing and wave-based computation with short wavelengths at GHz frequencies~\cite{Lenk2011, Yu2016}---without Joule heating due to the absence of charge transmission~\cite{Flebus2024}.
This charge neutrality, however, also makes their electrical control and detection challenging.
A promising way to mitigate this limitation is to exploit magnon chirality.
The additional internal chirality degree of freedom allows left- and right-handed modes to form distinct dynamical channels, which reduces backscattering and leads to more robust signal routing~\cite{xie2026generaltheorychiralsplitting}.
Furthermore, chiral magnons facilitate the coupling to spin-orbit effects such as the spin Hall effect~\cite{RevModPhys.87.1213, Chumak2012} and thereby open up possibilities for electrical generation and detection.

\begin{figure}
    \centering
    \includegraphics[width=1.0\linewidth]{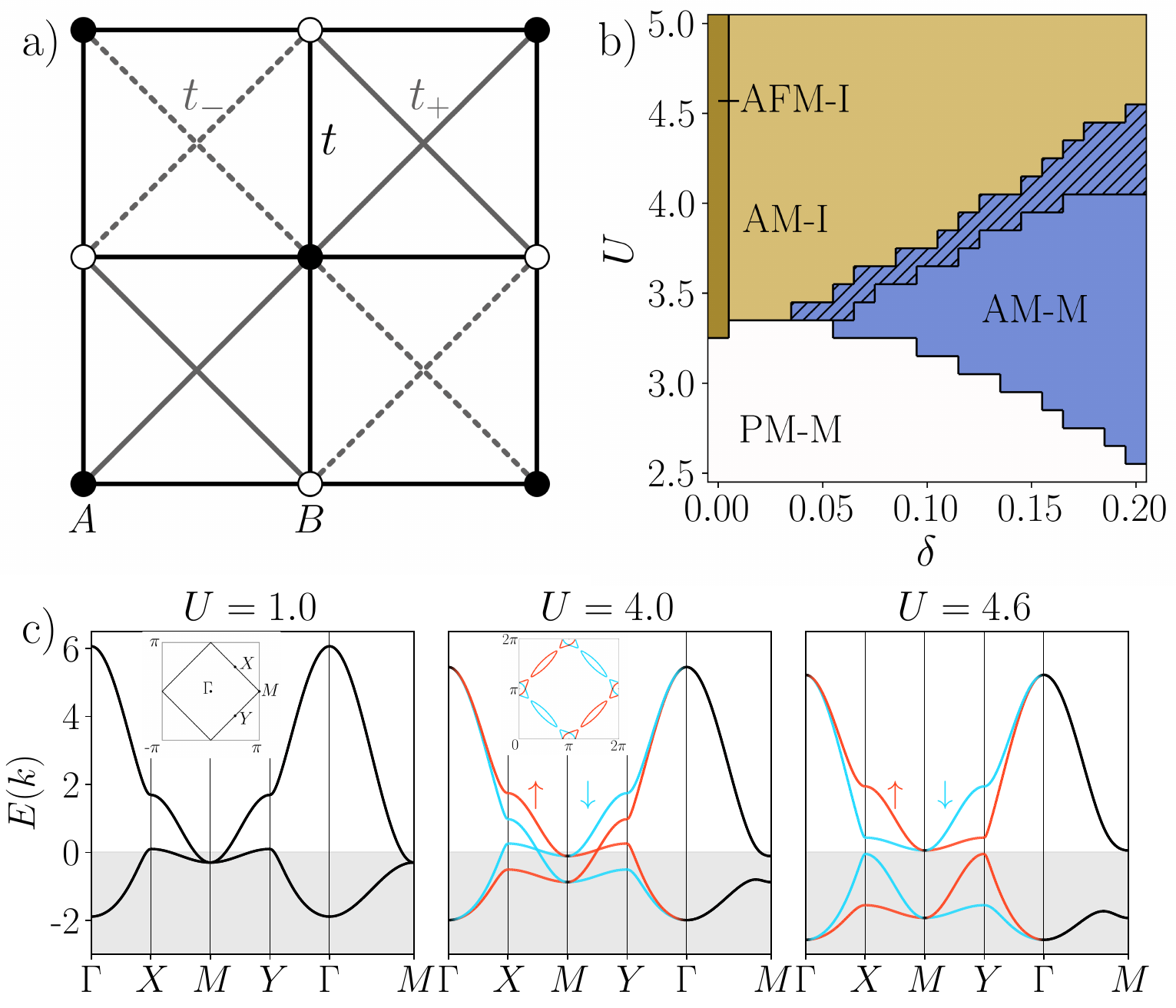}
    \caption{(a)~Checkerboard square lattice Hubbard model with anisotropic diagonal hoppings $t_\pm = t' \pm \delta$.
    (b)~Slave-boson mean-field phase diagram in the $U$--$\delta$ plane, comprising a paramagnetic metal (PM--M), an antiferromagnetic insulator (AFM--I), an altermagnetic metal (AM--M), and an altermagnetic insulator (AM--I).
    The hatched area marks an unstable region before the metal--insulator transition.
    (c)~Spin-polarized quasiparticle band structures for the PM--M, AM--M, and AM--I phases at $\delta=0.2$, plotted along the high-symmetry path shown in the inset.
    }
    \label{fig:lattice}
\end{figure}

A recent alternative source for realizing chiral magnons in a collinear magnet has been identified in altermagnets~\cite{PhysRevX.12.040501}. As opposed to spin-degenerate magnons of an antiferromagnet preserving $\mathcal{PT}$, where $\mathcal{P}$ denotes parity and $\mathcal{T}$ denotes time reversal symmetry, the altermagnet breaks $\mathcal{PT}$ so that the magnon branches energetically split into left-handed and right-handed chiral magnons~\cite{Smejkal2023, Liu2024}.
Altermagnetism finds its conceptual theoretical origin in the analysis of spin space groups (SSGs)~\cite{BrinkmanElliott1966,LitvinOpechowski1974}, allowing for a refined decomposition of symmetry generators as compared to magnetic space groups by intertwining space group symmetry with that of possible compensated magnetic order parameters.
From a simplified perspective, the prediction of altermagnetic materials currently follows a common scheme \cite{Roig2024, Guo2023}: Given a promising crystal,
(i)~determine its SSG to fall into the altermagnetic domain,
(ii)~calculate the ab initio structure under full consideration of spin-orbit coupling,
and (iii)~finally assume a magnetic mean-field to stabilize the desired altermagnetic electronic structure.
This electronic structure perspective is mirrored in many experiments on altermagnetic candidates, which primarily seek transport or spectroscopic signatures of spin-polarized bands \cite{Feng2022RuO2AHE, GonzalezBetancourt2023, Reimers2024, Krempasky2024}. The magnetic state that breaks time-reversal symmetry is then often treated only implicitly, despite being inseparable from the interaction-induced electronic correlations that stabilize and dress it.
Such an omission is consequential: magnetic fluctuations can decisively control both the stability and dynamical response of correlated magnetic states. \ch{RuO2} illustrates this point. First regarded as a primary altermagnetic candidate~\cite{Feng2022RuO2AHE, bose-ruo2, PhysRevLett.132.056701, Tschirner2023RuO2AHE, doi:10.1126/sciadv.adj4883}, it has become the subject of a polarizing debate, with recent work suggesting a highly fragile magnetic state or even the absence of magnetic order~\cite{PhysRevLett.133.176401,plouff,PhysRevLett.132.166702,keßler2024absencemagneticorderruo2,keßler2025moireassistedchargeinstabilityultrathin}. Beyond uncertainties in ab initio modeling, this underlines that, similar to issues associated with topological quantum chemistry~\cite{bradlyn,PhysRevB.96.121106}, {\it any classification and prediction of an electronic state of matter that oversimplifies the impact of Coulomb interactions necessitates a detailed subsequent many-body analysis beyond mean-field in order to reliably specify its nature from a perspective of microscopic theoretical modeling.}

In this Letter, we demonstrate that, even for a very simple yet generic model, incorporating electronic interactions alongside itinerancy profoundly affects the anisotropic lifetime and band structure of altermagnetic quasiparticles (chiral magnons). These chirality-dependent effects, which are readily missed in overly simplified weak-coupling approaches or conventional linear spin-wave theory, become most pronounced near the interaction-driven Metal-Insulator Transition (MIT). Using the Spin-Rotationally Invariant Slave-Boson method as a natural first step to go beyond such approaches, we develop a consistent description of correlated magnetic fluctuations across the full interaction range and establish a phase diagram, supported by Dynamical Mean Field Theory (DMFT). We then track the coupled evolution of magnonic and single-electron excitations across all relevant regimes from the dynamic slave-boson fluctuations.

\prlparagraph{Model \& Method}%
To enable a continuous interpolation between a conventional antiferromagnet and an altermagnetic state, we consider a minimal one-band Hubbard model on the checkerboard lattice, shown in Fig.~\ref{fig:lattice}(a). The Hamiltonian is given by
\begin{equation} \label{eq:hamiltonian_tt2d}
  \begin{aligned}
    H = &- t^\nodag \sum_{\langle i,j \rangle, \, \sigma}  \, c_{i,\sigma }^{\dagger}c^\nodag_{j,\sigma}
    - t_{\pm}  \sum_{\langle\!\langle i,j \rangle\!\rangle_{\pm}, \,\sigma} c_{i,\sigma }^{\dagger}c^\nodag_{j,\sigma}
    \\ &+ U \sum_{i} c_{i,\uparrow}^{\dagger}c^\nodag_{i,\uparrow} c_{i,\downarrow}^{\dagger}c^\nodag_{i,\downarrow} \ ,
  \end{aligned}
\end{equation}
where \(c_{i,\sigma}^{\dagger}\) (\(c_{i,\sigma}\)) creates (annihilates) an electron with spin \(\sigma\) at site \(i\).
The sum over \(\langle i,j \rangle\) includes nearest-neighbor bonds with hopping amplitude \(t\), while \(\langle\!\langle i,j \rangle\!\rangle_{\pm}\) refers to next-nearest-neighbor bonds along the two inequivalent diagonal directions of the checkerboard lattice. The corresponding hopping amplitudes are parametrized as \(t_{\pm} = t' \pm \delta\). In the following, we set \(t = 1\) as the unit of energy and fix \(t' = -0.3 \,t\), such that varying \(\delta\) directly tunes the hopping anisotropy.
For \(\delta = 0\), each lattice site preserves the full \(C_{4v}\) point-group symmetry of the checkerboard lattice, and staggered magnetic order corresponds to a conventional antiferromagnet.
A nonzero anisotropy \(\delta \neq 0\) reduces the local site symmetry to \(C_{2v}\), thereby allowing for a \(d\)-wave altermagnetic state within the two-sublattice unit cell~\cite{Roig2024,He2025,Khatua2025,Das2024,DelRe2025}.

With itinerant altermagnetism being studied predominantly within mean-field~\cite{Leeb2024,Giuli2025} and its stabilization in unbiased many-body frameworks only having been achieved recently~\cite{Durrnagel2025,Issing2026,Durrnagel2026}, a reliable description of quantum fluctuations is made indispensable for assessing stability of collective excitations.

Hence, we go beyond Hartree--Fock (HF) plus random-phase approximation (RPA) and incorporate local correlation effects of the Hubbard interaction $U$ already at the mean-field level.
To do so we employ the spin-rotationally invariant slave-boson (SB) formalism in the Kotliar--Ruckenstein (KR) scheme~\cite{Li1989b}, where the local Hilbert space is enlarged by auxiliary bosonic and pseudofermionic fields, while the physical subspace is enforced by local constraints via Lagrange-multipliers. The electron operator is then represented as $c^\dagger_{i\sigma} = \sum_{\varsigma} \underline{z}^\dagger_{i,\sigma\varsigma}\, f^\dagger_{i\varsigma}, $ where $f_{i\varsigma}$ denote the pseudofermions and $\underline{z}_{i,\sigma\varsigma}$ is the KR renormalization matrix depending on the SB fields [cf. \cref{eq:sb_representation}]. We then apply a static two-sublattice mean-field ansatz and determine the correlated ground state as the saddle point of the free energy with respect to the bosonic and Lagrange-multiplier fields. Collective excitations are obtained by formulating the theory as a path integral in radial gauge and expanding the effective action to Gaussian order around this saddle point. The resulting fluctuation propagator encodes the dynamical response, such that static and dynamical susceptibilities are computed from the corresponding slave-boson fluctuation correlators (cf. \cref{app:SRISB} for details and Ref.~\cite{Riegler2020} for a comprehensive overview).
Slave-boson approaches of this kind have proven highly effective in capturing dynamical signatures of strongly correlated phases --- such as effective masses, carrier lifetimes, spectral properties, and spin- and charge excitation spectra --- with direct connections to experimental probes including angle-resolved photoemission, neutron scattering, and resonant X-ray spectroscopy~\cite{Riegler2023,Klett2024}.

\begin{figure*}
    \centering
    \includegraphics[width=1\textwidth]{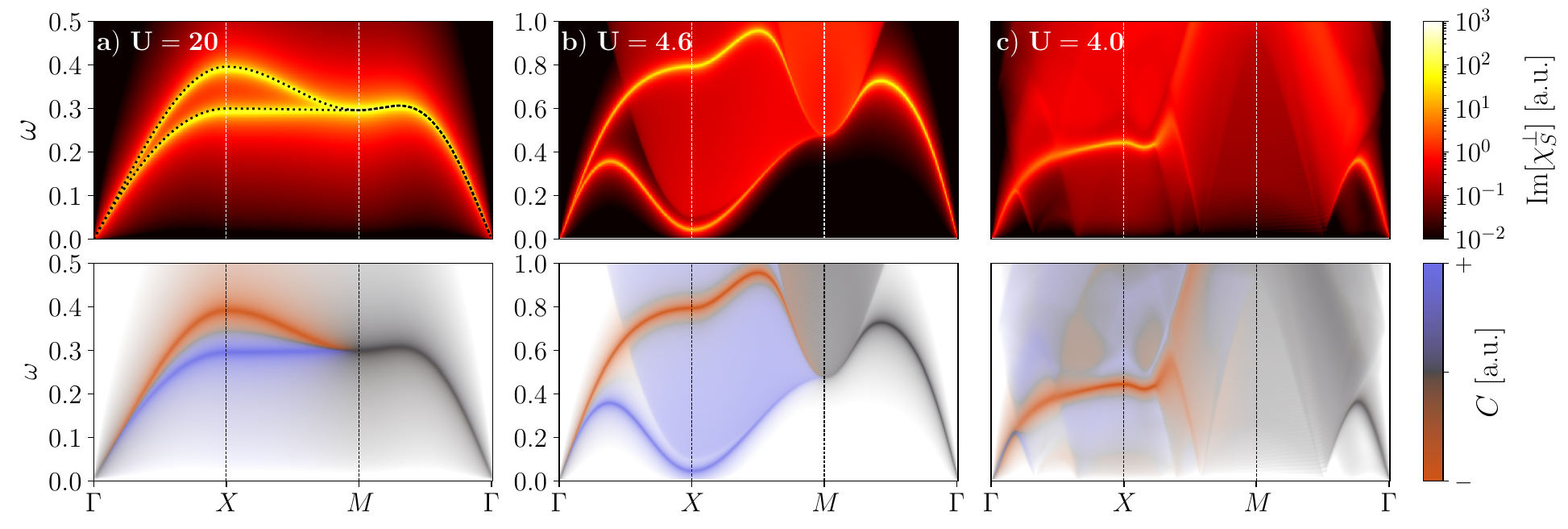}
    \caption{Top row: Imaginary part of the transverse spin susceptibility, $\mathrm{Im}\,[\chi_S^{\perp}]$, along the high-symmetry path $\Gamma$--$X$--$M$--$\Gamma$ for different values of $U$ at $\delta = 0.2$ corresponding to the AM--I (a)~far away from the MIT and (b)~close to the MIT, as well as (c)~the AM--M phase. The magnon dispersion obtained from linear spin-wave theory (LSWT) is shown by the black dotted lines in panel~(a). 
    Bottom row: Chirality-resolved magnon spectrum for the respective phases. Both magnons are degenerate along the nodal line $M$--$\Gamma$, such that their chiralities cancel each other, while each individual magnon branch as well as the associated continua remain chiral.
    }
    \label{fig:sus_chirality}
\end{figure*}

\prlparagraph{Phase diagram}%
The ground-state ($T=0.005$) phase diagram of the Hamiltonian [cf.~\cref{eq:hamiltonian_tt2d}] in the $U$--$\delta$ plane at half-filling obtained with SB mean-field theory is depicted in \cref{fig:lattice}(b). At large hopping anisotropy $\delta$, the paramagnetic metal (PM--M) and altermagnetic insulator (AM--I) are separated by an intermediate altermagnetic metal (AM--M). Upon decreasing $\delta$, this AM--M regime progressively narrows and disappears around $\delta \simeq 0.04$, below which the system undergoes a direct PM--M to AM--I transition. In the isotropic limit $t_+=t_-$, the altermagnetic splitting vanishes and the magnetic insulator reduces to the conventional antiferromagnetic insulator (AFM-I), yielding the expected direct PM--M to AFM--I transition.

The origin of this intermediate metallic regime lies in the anisotropic gap formation characteristic of altermagnetic order, governed by two distinct energy scales. First, the conventional AFM exchange splitting---set by the local magnetization---separates the relevant bands while preserving a residual double degeneracy protected by $\mathcal{P}\mathcal{T}$ symmetry. Second, the finite hopping anisotropy $\delta$ lifts this degeneracy and induces an additional, non-relativistic, momentum-dependent spin splitting. Crucially, in contrast to the exchange gap, this altermagnetic spin splitting is strongly momentum dependent and does not open uniformly across the Brillouin zone. As a result, upon the onset of magnetic order at $U_{c_1}$, only parts of the Fermi surface are gapped, while residual pockets remain (cf.~\cref{app:minimal_model}), yielding an indirect gap structure for $U_{c_1} < U < U_{c_2}$. At the critical interaction scale $U_{c_2}$, these pockets vanish and a fully insulating state is reached. The separation between $U_{c_1}$ and $U_{c_2}$, and consequently the width of the AM--M phase, grows with the anisotropy $\delta$. Increasing anisotropy effectively flattens the bands near the Fermi level, enhancing the energetic gain from interaction-induced spectral weight redistribution. Consequently, the interaction scale for the onset of magnetic order, $U_{c_1}$, is decreased. The MIT at $U_{c_2}$ is controlled by the indirect gap: stronger anisotropy enhances the momentum-dependent spin splitting and stabilizes residual Fermi pockets over a wider interaction range, shifting $U_{c_2}$ to larger values. A qualitatively similar intermediate AM--M regime between PM--M and AM--I has been reported in variational Monte Carlo studies of the Shastry--Sutherland lattice~\cite{Ferrari2024}.

We include a nearest-neighbor density interaction $V$ to assess the robustness of the AM--M and AM--I.
For $\delta = 0.2$, all phases remain stable up to at least $V = 1.0$, as we observe neither charge order nor divergent static susceptibilities. Furthermore, for small anisotropy, the direct PM--M to AM--I transition---and for $\delta \gtrsim 0.04$ the PM--M to AM--M transition---is first order, as evidenced by a discontinuous jump in the magnetization $m$ at the phase boundary (see~\cref{fig:maxwell_combined}). Additionally, the AM--M to AM--I transition is accompanied by a phase separated regime, as discussed in \cref{sec:phase_separation}, which also implies a first order phase transition. We corroborate these findings by DMFT calculations~\cite{Georges1996,Wallerberger2019}, which treat correlations by self-consistently mapping the lattice model onto a local impurity problem. At large anisotropy of \(\delta=0.2\), DMFT reproduces the succession of phase transitions from PM--M via AM--M to AM--I with very similar critical interaction values as in the SB calculations. Both transitions show clear hysteresis behavior, substantiating their first order nature. We note that the MIT at \(U_{c_2}\) is accompanied by a sharp increase of quasiparticle weight which otherwise decreases with increasing \(U\).

\begin{figure*}[t!]
  \includegraphics[width=0.5\linewidth]{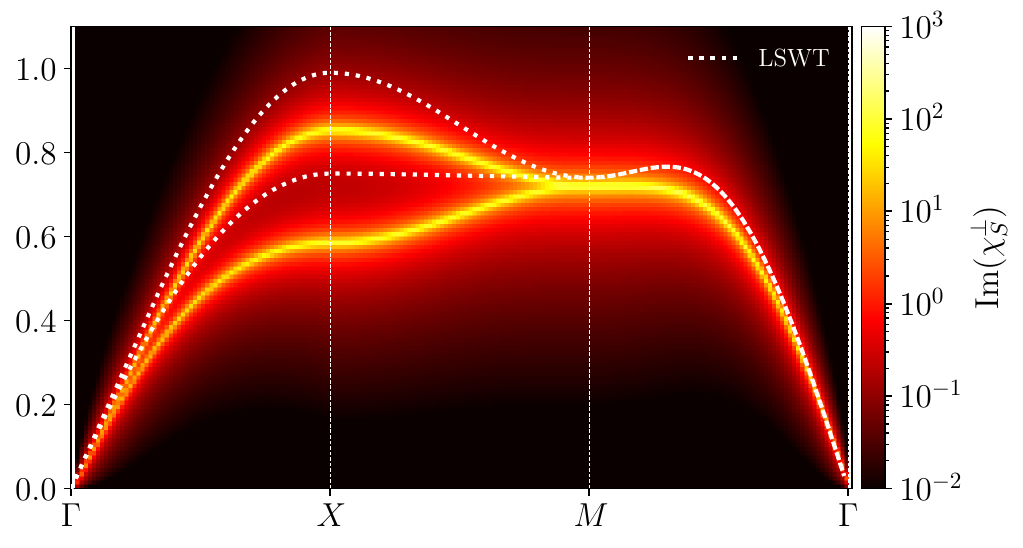}
  \caption{Magnon spectrum from LSWT for $S = 0.5,\ t = 1.0,\ t' = -0.3\, t,\ U = 8.0,\ \delta = 0.2$ overlaid onto the respective dynamical magnon spectrum for $U=8.0$ obtained with SBMFT.}
  \label{fig:LSWT}
\end{figure*}

\prlparagraph{Dynamical altermagnetic response}%
Going beyond the mean-field ground state, we study the dynamical spin response relevant for spectroscopic probes such as inelastic neutron scattering. This response is encoded in the chiral spin susceptibilities, which we evaluate from SB Gaussian fluctuations around the correlated saddle point. The corresponding equations are given in \cref{app:SRISB}. To quantify the magnon chirality, we define the measure
$\mathcal{C}(\bm q,\omega) = \Im\big(\chi_S^{+-}(\bm q,\omega) - \chi_S^{-+}(\bm q,\omega)\big)/2\Im\big(\chi_S^\perp(\bm q,\omega)\big)$.
As enforced by the underlying AM symmetry, the dynamical spectrum of AM phases contains chiral magnon branches~\cite{Smejkal2023}. In the strongly insulating regime [cf.~\cref{fig:sus_chirality}(a)], we indeed find sharp chiral magnon modes with nodal lines along the diagonals of the Brillouin zone, i.e., along $M$--$\Gamma$. This is consistent with the prediction from linear spin wave theory (LSWT; see \cref{app:LSWT} for a detailed derivation). However, agreement with LSWT is recovered only for $U\gg \Delta_{\mathrm{AM}}$, i.e., deep in the insulating regime. There, double occupancy is strongly depleted and electronic continua are shifted to high energies, so that the low-energy manifold is well described by an effective spin model. As a consequence, decreasing $U$ (and thereby effectively increasing double occupancy) 
leads to deviations in the SB magnon dispersion compared to LSWT~\cite{Garcia-Gaitan2025} (see~\cref{fig:LSWT}), where this double occupancy is projected out via the Schrieffer--Wolff transformation~\cite{Bravyi2011}.

\begin{figure}
    \centering
    \includegraphics[width=1.0\linewidth]{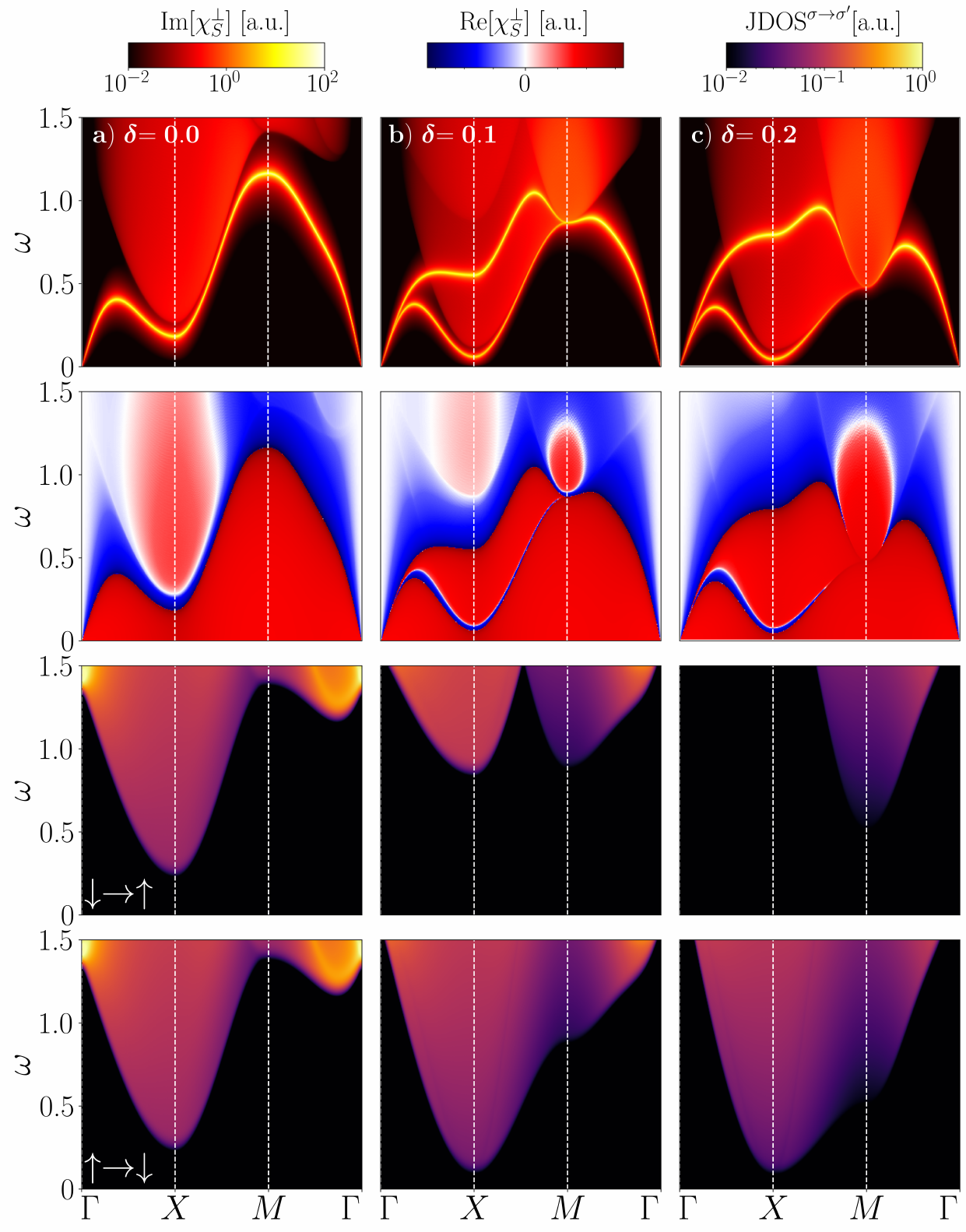}
    \caption{Imaginary (first row) and real parts (second row) of the transverse spin susceptibility, $\chi_s^\perp(\mathbf{Q},\omega)$, together with the chirality-resolved joint density of states (JDOS; bottom two rows) for spin-flip excitations, cf.~\cref{eq:jdos}, across the AFM--AM transition for increasing hopping anisotropy $\delta$, shown close to the metal--insulator transition at (a)~$U=3.5$, $\delta=0$, (b)~$U=3.9$, $\delta=0.1$, and (c)~$U=4.6$, $\delta=0.2$. Upon increasing $\delta$, the interaction between magnons and the Stoner continuum evolves in an anisotropic fashion, resulting in momentum-dependent magnon stability.
    }
    \label{fig:afm-am}
\end{figure}

\prlparagraph{Lifetime anisotropy}%
Close to the MIT as $U$ approaches $U_{c_2}$ from above, the magnon branches are rescaled due to the electrons' finite quasiparticle weight. Additionally, they are deformed by pronounced low-energy Stoner excitations, as visualized in \cref{fig:sus_chirality}(b).
As a consequence, the magnon response can no longer be understood as that of an isolated collective mode, but must be considered together with the nearby excitation continuum. 
The system's altermagnetic nature dictates that the Stoner continuum affects different regions of the Brillouin zone in a strongly anisotropic way:
At $M$, both magnon branches exhibit significant deformation and touch the continuum.
They thus acquire finite lifetime by decaying into particle--hole excitations, which leads to a loss of spectral weight, i.e., Landau damping~\cite{Costa2025, Wang2024, Gaveau1995}. 
This decay is constrained by chirality, since magnons can only couple to continuum states in the same chiral channel.
Mixing between opposite chiralities is forbidden by the remaining $U(1)$ symmetry in the AM state:
Spin diagonal structure of the electronic propagators renders chirality a good quantum number in presence of magnon--electron as well as multi--magnon scattering processes~\cite{Eto2025}, which results in full chirality of all modes in \cref{fig:sus_chirality}. 
On the nodal line, where the continuum consists of both chiralities, the underlying AM symmetry enforces both magnon branches to decay simultaneously (at $M$), but they remain separate.
In contrast, the low-energy continuum at $X$ is strongly chirality-polarized, such that only the positive chirality magnon is renormalized and pushed outside the continuum to lower frequencies, while the negative chirality magnon is unaffected.
Notably, Landau damping is absent at the $X$ point. Owing to symmetry, the $Y$ point exhibits behavior analogous to that of the $X$ point, but for the opposite chirality.

To better understand this feature, we analyze the magnon spectrum for different values of $(\delta,U) = (0.0, 3.5), (0.1, 3.9)$ and $(0.2, 4.6)$ in \cref{fig:afm-am}, all close to the MIT.
The special case $\delta = 0.0$ corresponds to the AFM limit, where no anisotropy is present.
Here, the two magnon branches are degenerate throughout the Brillouin zone. The sharp magnon excitations strongly repel the continua and are therefore damped at neither the  $X$ nor the $M$ point.
Whether magnon and continuum touch is determined by a sign flip in the real part of $\chi_S^\perp$ (cf.~second row of \cref{fig:afm-am}), that separates the sharp mode from the continuum.
This gap shrinks and eventually vanishes at $M$ as $\delta$ is increased, leading to substantial Landau damping even within the insulating phase [see~\cref{fig:afm-am}(c)].
The basic mechanism behind this level repulsion is the coupling of a discrete level, e.g., magnon, with a nearby particle-hole continuum~\cite{Gaveau1995}.
Whether the branch is repelled or dampened is controlled by the effective strength of the continuum~\cite{Verresen2019, Wang2024, Plumb2016}:
the stronger the continuum, the more pronounced becomes the level repulsion and eventually leads to a full repulsion of the magnon branch out of the continuum such that it retains its sharp spectral features.
If it is too small, the mode experiences only slight deformation but enters the continuum at some point and undergoes Landau damping.
We directly attribute these different behaviors to the change of the Stoner continua as a function of $\delta$, which we quantify by means of the joint density of states (JDOS) [cf.~\cref{eq:jdos}].
Inspecting the last two rows of \cref{fig:afm-am}, we see that the Stoner continuum at $X$ remains largely unaffected by the anisotropy $\delta$, which only leads to a splitting of the two chiral continua, while largely retaining their shape and strength.
The emergent spin splitting with increasing $\delta$, however, leads to a lowering of the excitation continuum at $\boldsymbol{Q} = M$ and replaces the strong continuum close to $\Gamma$ in the AFM as the minimum.
This continuum however, is weak as attested by a small JDOS, since the curvature of the bands connected by the nesting vector $M$ does not match (cf.~\cref{fig:excitations}).
Consequently, the level repulsion is marginal, and the continuum at $M$ eventually touches the magnon branches as it moves down in energy, leading to a substantial loss of spectral weight around the nodal line.
This, in turn, makes the degenerate magnon modes decay with a finite lifetime.
\emph{En passant}, we note that the deformation of the magnon dispersion is expected to significantly alter the nature of multi-magnon decay~\cite{Eto2025}.

\prlparagraph{Magnon chirality in the AM metal}%
The level repulsion of the lower chiral branch at $X$ is lost once the system enters the metallic phase, where the magnon energy overlaps with continuum states in the same chiral sector and Landau damping sets in, submitting the collective mode to decay into particle--hole excitations. Spectroscopically, this is seen as a broadening of the magnon peak, a loss of spectral weight, and ultimately the loss of a well-defined pole in the transverse susceptibility in \cref{fig:sus_chirality}(c). Interestingly, we also observe that the magnon with opposite chirality to the continuum substantially loses spectral weight in addition to its band renormalization, but retains its symmetry-protected chirality. Both findings are a direct consequence of the suppressed chirality of the continuum in the AM--M and the loss of a well-defined edge of the continuum in the presence of residual Fermi surfaces in the AM--M due to the indirect gap. Away from $\Gamma$, the transfer of spectral weight from the magnon branches by Landau damping is inevitable even at zero energy, leading to finite lifetimes of the quasiparticle modes. Consequently only a single branch with well-defined chirality and substantial lifetime remains in the magnon spectrum. This renders altermagnetic metals a promising platform not only for spintronic applications with electronic quasi-particles, but also for magnonic applications with elementary spin excitations~\cite{Beida2025}.

\prlparagraph{Conclusion \& Outlook}%
In this Letter, we present the first consistent study of magnons in altermagnetic system covering all energy scales of the AM gap from the Mott limit all the way down to the metallic regime in a single framework using SBMFT.
This bridges the current gap between preceding studies in the strong~\cite{Eto2025} and weak coupling limit~\cite{Costa2025}.
Compared to RPA, the present approach yields a more consistent picture across the full interaction regime by incorporating fluctuation corrections already at the level of the reference state. 
This turns out to be crucial for a concise prediction of dynamical correlation functions, that require a consistent treatment of the collective response and the magnetic parent state to yield meaningful results already on the MF+RPA level (cf.~\cref{app:RPA}). 

We find two sharp magnon modes of both chiralities on the insulating side of the MIT, which become strongly renormalized by the electron--hole continua close by.
In contrast, only one chiral magnon branch survives Landau damping in the metal. 
Our results show that the altermagnetic MIT provides access to two complementary magnonic functionalities.
On the weakly insulating side, chiral magnons remain sharp despite strong renormalization and exhibit enhanced correlation induced splitting close to the transition, making this regime more attractive as the magnons become more easily selectively addressable as for smaller splitting. In the metallic phase, by contrast, chirality- and momentum-selective damping suppresses competing channels while one chiral branch remains sharp, pointing instead to a dissipative magnonic regime in which selective damping may be harnessed for fast magnetization switching~\cite{Beida2025, Wang2019, Choi2025}.
The vicinity of the transition thus emerges as a crossover between coherent and dissipative chiral magnonics.

\prlparagraph{Acknowledgments}%
We thank P. Wölfle and A. Maity for fruitful discussions. 
This research was funded by the Deutsche Forschungsgemeinschaft (DFG, German Research Foundation) – Project-ID 258499086 – SFB 1170; through the Würzburg-Dresden Cluster of Excellence on Complexity, Topology, and Dynamics in Quantum Materials (ctd.qmat) – Project-ID 390858490 – EXC 2147; and through the Research Unit QUAST – Project-ID 449872909 – FOR 5249. MD acknowledges support from the Studienstiftung des deutschen Volkes.

\let\oldaddcontentsline\addcontentsline
\renewcommand{\addcontentsline}[3]{}
\bibliography{bibliography}
\let\addcontentsline\oldaddcontentsline

\supplement{Supplemental Material:\\
Altermagnons at the metal-insulator transition}

\tableofcontents

\section{Spin-Rotationally Invariant Kotliar-Ruckenstein Slave-Bosons}
\label{app:SRISB}

\subsection{Formalism}
To incorporate local correlation effects beyond Hartree--Fock, we employ the spin-rotation-invariant Kotliar--Ruckenstein slave-boson formalism~\cite{{Kotliar1986},{Fresard1992}}. Therefore, we introduce six slave bosons $b^\dagger \in \{e,p,d\}$, with $p$ being a $2\times2$ matrix, and two "pseudofermions" $f^\dagger_\uparrow, f^\dagger_\downarrow$ to label the four states of a single site with two spin-orbitals plus a total of five constraints
{
    \begin{subequations}
        \begin{align} \label{eq:SRISB_rep}
        \ket{0} \to e^\dagger \ket{\textrm{vac}} \quad
        \ket{\sigma} \to  \sum_\varsigma p^\dagger_{\sigma\varsigma} f^\dagger_{\varsigma}  \ket{\textrm{vac}} \quad
        \ket{2} \to  d^\dagger f^\dagger_{\uparrow} f^\dagger_{\downarrow} &\ket{\textrm{vac}} \\
        Q_1 = e^\dagger e^\nodag + \Tr \, p^\dagger p^\nodag + d^\dagger d^\nodag - 1 = 0 \label{eq:completeness}\\
        Q_2^\alpha = 
        \Tr \, p^\dagger \tau^{\alpha} p^\nodag + 2 d^\dagger d^\nodag \delta_{\alpha 0} - \sum_{\varsigma\varsigma'} f^\dagger_{\varsigma} \tau^{\alpha}_{\varsigma\varsigma'} f^\nodag_{\varsigma'} = 0 \label{eq:fillingconstraint}
        \end{align}
    \end{subequations}
}
where $\tau^\alpha$ with $\alpha = 0,1,2,3$ are the usual four Pauli-matrices. The physical electron operator is represented in the enlarged Hilbert space as
\begin{equation}
\label{eq:sb_representation}
c^\dagger_{i\sigma}
\stackrel{\text{constraints}}{=}
\sum_{\varsigma}
\sqrt{2}\,L_i(b)\,
\Big(
p^\dagger_{i,\sigma\varsigma}\, e_i
+ \, d^\dagger_i\, \tilde p_{i,\sigma\varsigma}
\Big)\,
R_i(b)\,
f^\dagger_{i\varsigma}\, ,
\end{equation}
where $e_i$, $d_i$, and $p_{i,\sigma\varsigma}$ denote the slave-boson fields, and $L_i(b)$ and $R_i(b)$ are the standard Kotliar--Ruckenstein renormalization operators. They reduce to unity on the physical subspace, but already at saddle-point level generate the familiar quasiparticle-weight renormalization of the hopping amplitudes. Here $\tilde p_i$ denotes the time-reversed partner of the $p_i$ matrix.

To formulate the theory in a way that allows for both a saddle-point treatment and the calculation of collective modes, we rewrite the problem as a path integral in radial gauge, where the phases of all slave-boson fields, except the $d_i$ field, are gauged away, such that $d_i=d_{1,i}+id_{2,i}$ remains the only complex bosonic field, while the remaining bosons are real-valued.  In this formulation, the physical content of the method is encoded in the saddle-point values of the bosonic fields, while fluctuations around that saddle point describe collective excitations. The partition function $\mathcal{Z}$ is given by
\begin{equation}
\begin{aligned}
\mathcal{Z}
&= \int \mathcal{D}[\bm{\phi}]\, e^{-S_{\mathrm{eff}}[\bm{\phi}]}\, ,\\
S_{\mathrm{eff}}[\bm{\phi}]
&=
S^{(0)}[\bar{\bm{\phi}}]
+ \sum_{q}\sum_{a,b}
\delta \bm{\phi}^{\dagger}_{\text{-}q,a}\,
\underline{D}^{-1}_{ab}(q)\,
\delta \bm{\phi}_{q,b}\, ,
\end{aligned}
\end{equation}
where $\bm{\phi}$ is the vector of all real slave-boson and Lagrange-multiplier fields, $\bar{\bm{\phi}}$ denotes their saddle-point values, and $\delta\bm{\phi}=\bm{\phi}-\bar{\bm{\phi}}$ are the associated fluctuation fields. Furthermore $q=(\mathbf{q},\omega)$, and
\begin{equation}
\underline{D}^{-1}_{ab}(q)\equiv
\frac{\delta^2 S_{\mathrm{eff}}}{\delta \bm{\phi}_a\, \delta \bm{\phi}'_b}
\end{equation}
is the inverse propagator of Gaussian fluctuations (fluctuation matrix). Within this formalism the constraints, ensuring completeness [\cref{eq:completeness}] and consistency between fermionic and bosonic occupation [\cref{eq:fillingconstraint}], can easily be enforced with the help of projectors. The effective action $S_{\mathrm{eff}}$ consists of the effective Lagrangian, which separates into a quadratic fermionic and a purely bosonic part. 
\begin{align}
    \mathcal{L}_{\text{eff}} [f, \phi] =& \mathcal{L}_\text{F} + \mathcal{L}_\text{B} = \sum_{\boldsymbol{k}_1, \boldsymbol{k}_2} \boldsymbol{f}_{\boldsymbol{k}_1}^\dagger (\partial_\tau + \underline{H}_{\boldsymbol{k}_1, \boldsymbol{k}_2}[\phi]) \boldsymbol{f}_{\boldsymbol{k}_2} \\ &+ \sum_i [d_i^{*}(\partial_\tau + U) d_i + i\alpha_i ( e_i^{2} + p_{0,i}^{2} + \boldsymbol{p}_i^{2} + d_i^* d_i -1) -  i \beta_{0,i}(p_{0,i}^{2} + \boldsymbol{p}_i^{2} + 2d_i^{*}d_i) - i \boldsymbol{\beta}_i \cdot 2 p_{0,i}\boldsymbol{p}_i]
\end{align} 
with 
    
\begin{align}
    \underline{H}_{\boldsymbol{k}_1, \boldsymbol{k}_2} [\phi] &= - \mu_0 \underline{1\!\!1} _2 \delta_{\boldsymbol{k}_1,\boldsymbol{k}_2} + \sqrt{\frac{1}{N}} (\underline{\beta})^{T}_{\boldsymbol{k}_1 - \boldsymbol{k}_2} + \frac{1}{N} \sum_{\boldsymbol{k}} (\underline{z}^\dagger)^{T}_{\boldsymbol{k} - \boldsymbol{k}_1} \underline{\mathcal{H}}_{\boldsymbol{k}} (\underline{z})^{T}_{\boldsymbol{k} - \boldsymbol{k}_2} \\ 
    \underline{\beta} &= \sum_{\nu=0}^{3} \beta_\nu \underline{\tau}^{\nu}
\end{align}
where $\underline{\mathcal H}_{\boldsymbol{k}}$ is the bare hopping matrix of the underlying two-sublattice Hubbard model and $\underline{z}$ is the slave-boson quasiparticle-renormalization matrix. The latter renormalizes the hopping amplitudes and encodes the suppression of coherent charge motion by local correlations already at saddle-point level.

To describe the ordered AM phases, we adopt a static two-sublattice saddle-point ansatz. All SB fields and Lagrange multipliers are taken to be time independent, but may assume different values on the two sublattices $A$ and $B$. This reduces the pseudofermionic sector to a quadratic problem in the enlarged sublattice-spin space and yields the correlated quasiparticle band structure of the ordered state. The saddle-point solution is obtained by extremizing the corresponding free energy with respect to the bosonic and Lagrange-multiplier fields. 

Dynamical response functions are now obtained by expanding the corresponding observables to linear order in the fluctuation fields and contracting them with the fluctuation propagator $D$. In the slave-boson representation, the spin-density operator is given by
\begin{equation}
    \mathbf{S}_i = p_{0,i}\,\mathbf{p}_i,
    \qquad
    \mathbf{p}_i=(p_{1,i},-p_{2,i},p_{3,i})^T .
\end{equation}
Accordingly, the spin susceptibility is defined as
\begin{equation}
    \chi_S^{\mu \nu}(q,\omega)
    = \left\langle T\,S^\mu_{-q}S^\nu_q\right\rangle(\omega)
\end{equation}
which, at Gaussian level, is obtained as a weighted combination of matrix elements of the fluctuation propagator. In the two-sublattice parametrization this yields the site-resolved susceptibility
\begin{equation}
    \chi_{S,ab}^{\mu \nu} = \bar p_{0,a} \bar p_{0,b} \cdot D_{p_{\mu,a}, p_{\nu, b}} + \bar p_{\mu,a}\bar p_{0,b} \cdot D_{p_{0,a}, p_{\nu,b}} + \bar p_{0,a} \bar p_{\nu,b} \cdot D_{p_{\mu,a},p_{0,b}} + \bar p_{\mu,a} \bar p_{\nu,b} \cdot D_{p_{0,a},p_{0,b}} 
\end{equation}
where $a,b \in \{A,B\}$. Hence, the dynamical spin susceptibility is fully determined by the propagators of the relevant slave-boson fluctuation fields. The first term describes direct correlations between spin-like fluctuations $\delta p_{\mu}$, while the remaining terms account for the mixing between amplitude fluctuations of $p_0$ and the spin-carrying bosons $p_{\mu}$. In this way, the collective spin response naturally incorporates the coupling between spin and local charge configurations already at Gaussian level. The full susceptibility is then calculated by summing over all site pairs with the corresponding sublattice phase factor,
\begin{equation}
    \chi_S^{\mu \nu} = \frac{1}{N_c} \sum_{ab} e^{i \boldsymbol{Q} (\boldsymbol{r}_a - \boldsymbol{r}_b)} \cdot \chi_{S,ab}^{\mu \nu}
\end{equation}
To resolve the chiral response, it is convenient to transform from the Cartesian basis $\mu,\nu\in\{x,y\}$ to the circular basis defined by
\begin{equation}
    S^\pm = S^x \pm i S^y .
\end{equation}
The susceptibilities in the chiral basis as well as the transverse spin susceptibility are then obtained from the Cartesian components $\chi_S^{\mu\nu}$ according to
\begin{align}
    \chi_S^{+-}(q,\omega)
    &= \chi_S^{xx}(q,\omega)
    - i\!\left[\chi_S^{xy}(q,\omega)-\chi_S^{yx}(q,\omega)\right]
    + \chi_S^{yy}(q,\omega) \\
    \chi_S^{-+}(q,\omega)
    &= \chi_S^{xx}(q,\omega)
    + i\!\left[\chi_S^{xy}(q,\omega)-\chi_S^{yx}(q,\omega)\right]
    + \chi_S^{yy}(q,\omega) \\
    \chi_S^{\perp}(q,\omega) &= \frac{1}{2} [\chi_S^{xx}(q,\omega) + \chi_S^{yy}(q,\omega)] 
\end{align} 

\subsection{Analysis of the unstable regime before the metal-insulator transition}
\label{sec:phase_separation}

\begin{figure}[t]
    \centering
    \begin{minipage}[t]{0.48\linewidth}
        \centering
        \includegraphics[width=\linewidth]{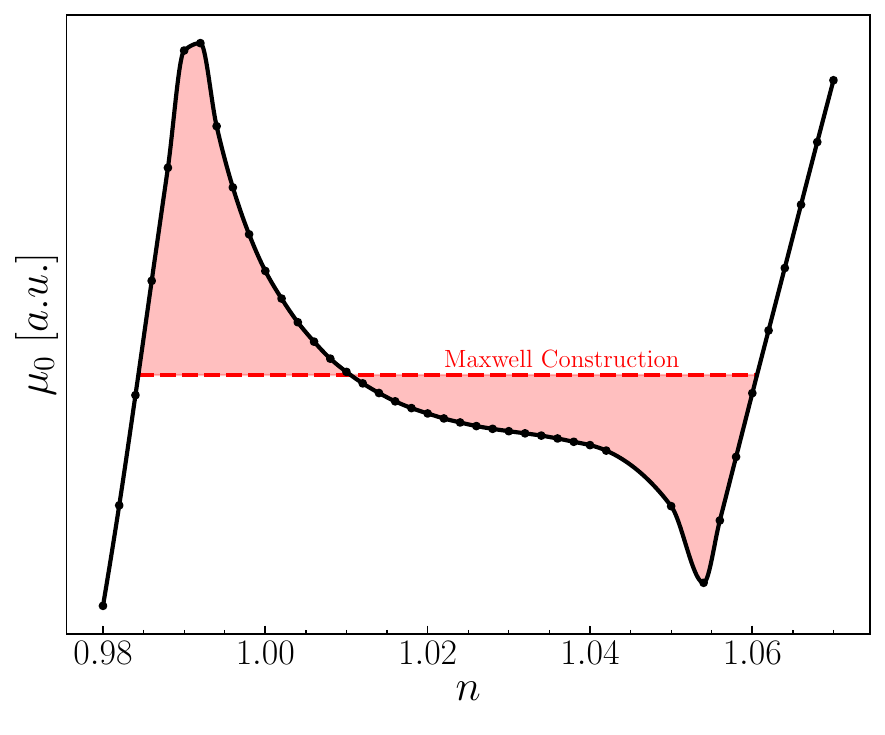}
    \end{minipage}
    \hfill
    \begin{minipage}[t]{0.48\linewidth}
        \centering
        \includegraphics[width=\linewidth]{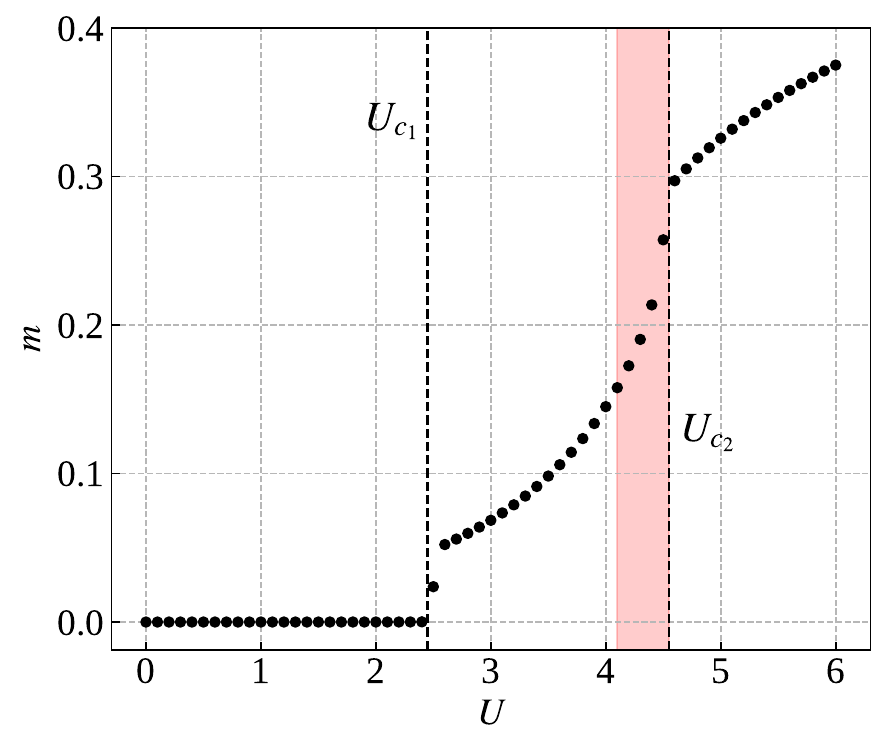}
    \end{minipage}
    
    \caption{Left: Chemical potential $\mu_0$ versus filling $n$ in the unstable regime before the MIT at $U=4.4$ and $\delta=0.2$. The region with negative slope $\partial \mu_0/\partial n < 0$ signals negative compressibility and a tendency toward macroscopic phase separation. The dashed line indicates the Maxwell construction used to determine the coexistence window between an altermagnetic and a paramagnetic state. Right: Magnetization $m$ over $U$ for $\delta=0.2$ with PM--M to AM--M ($U_{c_1}$) and the AM--M to AM--I ($U_{c_2}$) marked by the vertical dashed lines. The unstable region is marked in red.}
    \label{fig:maxwell_combined}
\end{figure}

A pronounced unstable region appears before the metal--insulator transition (MIT), corresponding to the hatched area in Fig.~\ref{fig:lattice}(b). Within this regime, the charge susceptibility $\chi_c$ diverges at $\Gamma$, while the longitudinal spin susceptibility $\chi_S^{||}$ diverges at the $M$ point. As discussed previously in Refs.~\cite{Issing2026,Seufert2021}, such simultaneous instabilities are a characteristic indication of phase separation. In addition, precisely at the transition, the transverse spin susceptibility $\chi_s^{\perp}$ exhibits a divergence at $(\pi/2,\pi/2)$. In contrast to the divergences in $\chi_c$ and $\chi_S^{||}$, which disappear at the transition, this transverse instability persists over a narrow interval on the insulating side of the phase boundary.
To test the interpretation in terms of phase separation more directly, we consider the point $\delta=0.2$ and $U=4.4$ inside the unstable region and perform a filling sweep to extract the chemical potential $\mu_0(n)$. As shown in Fig.~\ref{fig:maxwell_combined} left hand side, the resulting curve contains a region with negative slope, $\partial \mu_0/\partial n < 0$, corresponding to negative electronic compressibility. This behavior is a standard signature of phase separation and is therefore fully consistent with the fluctuation analysis. The dashed line in the same figure indicates the Maxwell construction used to determine the coexistence window between the competing phases.
The mean-field solution in this regime points to a coexistence between an altermagnetic metal (AM--M) and a paramagnetic metal (PM--M). More specifically, for fillings in the range $n=0.98$--$1.05$, the mean-field state is AM--M, whereas for $n>1.05$ the PM--M becomes energetically favorable. This crossover coincides with the point where $\mu_0(n)$ starts to increase again. Furthermore, at $n=0.99$ the Fermi-surface topology changes through the appearance of an additional pocket around $(\pi,0)$, which is also the filling where the instability in the charge susceptibility sets in. Taken together, these observations indicate that the phase-separated regime consists of coexisting AM--M and PM--M states.

\subsection{Relevant excitations}

\begin{figure}
    \centering
    \includegraphics[width=0.5\linewidth]{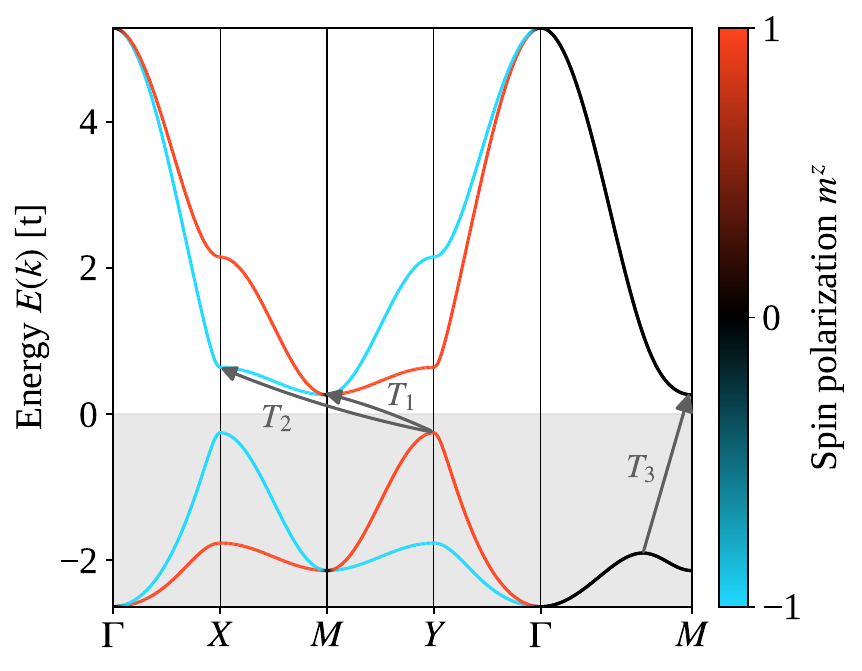}
    \caption{Relevant Transitions $T_{1-3}$ for the MIT as we increase $\delta=0-0.2$. Note that $T_2$ here connects $(\pi/2,-\pi/2) \rightarrow (\pi/2,\pi/2)$ with $\boldsymbol{Q} = (0,\pi)$, which is not shown along the high-symmetry path in \cref{fig:sus_chirality,fig:afm-am}. However, due to symmetry, this transition is equivalent to the transition $(-\pi/2,\pi/2) \rightarrow (\pi/2,\pi/2)$ with $\boldsymbol{Q} = (\pi, 0)$.}
    \label{fig:excitations}
\end{figure}

To illustrate the excitations that control the behavior across the AFM--AM transition, the most relevant transitions are marked by arrows in the band structure shown in Fig.~\ref{fig:excitations}. Among the various possible particle-hole processes, the three transitions labeled $T_{1}$--$T_{3}$ play a particularly important role in the discussion of the magnon spectrum across the transition. In particular, they provide the minimal set of processes needed to understand the momentum-dependent evolution of the collective spin response across the AFM--AM transition.

As a measure of the continuum strength we introduce the joint density of states (JDOS) given by~\cite{OLeary2002, Orapunt2008}
\begin{equation}
    \label{eq:jdos}
    \text{JDOS}^{\sigma \rightarrow \sigma'} (\boldsymbol{Q}, \omega) = \frac{1}{N_k} \sum_{\boldsymbol{k}, n, m} f\big(E_{n,\sigma}(\boldsymbol{k})\big) \,
    \Big(1-f\big(E_{m,\sigma'}(\boldsymbol{k}+\boldsymbol{Q})\big)\Big) \, \delta \big(\omega - E_{m,\sigma'}(\boldsymbol{k}+\boldsymbol{Q}) + E_{n,\sigma}(\boldsymbol{k}) \big) \, ,
\end{equation}
where $f(E)$ is the Fermi distribution. The quantity $\text{JDOS}^{\sigma \rightarrow \sigma'}(\boldsymbol{Q},\omega)$ measures the available phase space for particle-hole excitations with momentum transfer $\boldsymbol{Q}$ and energy transfer $\omega$. Since it is closely related to the bare transverse spin susceptibility, $\Im \chi_0^{\pm}(\boldsymbol{Q},\omega)$, it provides a useful indicator of the strength of the excitation continuum.

\clearpage
\section{Quasi-particle band structure analysis}
\label{app:minimal_model}

The minimal description of the effective model described in Eq.~\ref{eq:hamiltonian_tt2d}, can be written as 
\begin{equation}
H_\sigma(\mathbf{k})=
\begin{pmatrix}
\xi_{A\sigma}(\mathbf{k}) & \gamma(\mathbf{k})\\
\gamma(\mathbf{k}) & \xi_{B\sigma}(\mathbf{k})
\end{pmatrix},
\qquad \sigma=\pm 1 ,
\label{eqn:H_MF}
\end{equation}
with
\begin{align}
\gamma(\mathbf{k})&=-2t\bigl(\cos k_x+\cos k_y\bigr),\\
\xi_{A\sigma}(\mathbf{k})
&=
-2(t'+\delta)\cos(k_x+k_y)
-2(t'-\delta)\cos(k_x-k_y)
+\sigma M-\mu,\\
\xi_{B\sigma}(\mathbf{k})
&=
-2(t'-\delta)\cos(k_x+k_y)
-2(t'+\delta)\cos(k_x-k_y)
-\sigma M-\mu.
\end{align}
 eigenvalues are
\begin{equation}
E_{\sigma,\pm}(\mathbf{k})
=
\frac{\xi_{A\sigma}(\mathbf{k})+\xi_{B\sigma}(\mathbf{k})}{2}
\pm
\sqrt{
\left[
\frac{\xi_{A\sigma}(\mathbf{k})-\xi_{B\sigma}(\mathbf{k})}{2}
\right]^2
+\gamma^2(\mathbf{k})
}.
\end{equation}

\noindent
Hamiltonian in sublattice space,
\begin{equation}
H_\sigma(\mathbf{k})
=
\varepsilon_0(\mathbf{k})\,\tau_0
+
\gamma(\mathbf{k})\,\tau_x
+
m_\sigma(\mathbf{k})\,\tau_z ,
\end{equation}
where
\begin{align}
\varepsilon_0(\mathbf{k})
&=
\frac{\xi_{A\sigma}(\mathbf{k})+\xi_{B\sigma}(\mathbf{k})}{2}
=
-2t'\!\left[\cos(k_x+k_y)+\cos(k_x-k_y)\right]-\mu ,
\\
m_\sigma(\mathbf{k})
&=
\frac{\xi_{A\sigma}(\mathbf{k})-\xi_{B\sigma}(\mathbf{k})}{2}
=
\sigma M
-\,
2\delta\!\left[\cos(k_x+k_y)-\cos(k_x-k_y)\right] .
\end{align}

\begin{align}
\varepsilon_0(\mathbf{k})&=-4t'\cos k_x\cos k_y-\mu,\\
m_\sigma(\mathbf{k})&=\sigma M+4\delta \sin k_x\sin k_y . \label{eq:m_sigma}
\end{align}

\begin{equation}
E_{\sigma,\pm}(\mathbf{k})
=
\varepsilon_0(\mathbf{k})
\pm
\sqrt{\gamma^2(\mathbf{k})+m_\sigma^2(\mathbf{k})}.
\label{eq:minimal_bands_compact}
\end{equation}

\noindent
This representation makes the distinct roles of the different terms transparent. The scalar term
$\varepsilon_0(\mathbf{k})\tau_0$ shifts both bands equally and is fully symmetric under the lattice point group. The inter-sublattice hopping $\gamma(\mathbf{k})\tau_x$ hybridizes the two sublattices and likewise transforms trivially. By contrast, the magnetic mass $m_\sigma(\mathbf{k})\tau_z$ contains two qualitatively different contributions: a momentum-independent staggered field $\sigma M$, corresponding to the conventional antiferromagnetic order, and a momentum-dependent term proportional to $\delta \sin k_x\sin k_y$, which changes sign in momentum space and thus encodes the altermagnetic anisotropy. In the antiferromagnetic limit $\delta=0$, Eq.~\eqref{eq:minimal_bands_compact} reduces to
\begin{equation}
E_{\sigma,\pm}^{\mathrm{AFM}}(\mathbf{k})
=
\varepsilon_0(\mathbf{k})
\pm
\sqrt{\gamma^2(\mathbf{k})+M^2}.
\label{eq:AFM_bands}
\end{equation}

Hence the ordered state is spin degenerate and the magnetic order enters solely through the momentum-independent mass $M$. Hence the direct gap $\Delta E^{\mathrm{AFM}}=2\sqrt{\gamma^2(\mathbf{k})+M^2} \geq 2|M|$ $\forall ~\mathbf{k}$. In this sense, the AFM gap opens uniformly in $k$ space. At half filling, the onset of this magnetic gap leads to the AFM state being insulating in the parameter regime considered here. For finite anisotropy $\delta\neq0$, however, \cref{eq:m_sigma} becomes $k$-dependent and the gap opens anisotropically, such that it can no longer gap the entire Brillouin zone uniformly and an intermediate metal can emerge.

\section{Linear Spin Wave Theory} 
\label{app:LSWT}

\subsection{Effective spin Hamiltonian}

At half filling and $U\gg t,t'$, the Hubbard model maps (to second order in hopping) to
\begin{align}
H
= J_1\sum_{\langle ij\rangle}\bS_i\cdot\bS_j
+\sum_{\langle\!\langle ij\rangle\!\rangle}J_{ij}\,\bS_i\cdot\bS_j,
\qquad
J_1=\frac{4t^2}{U},\quad
J_{ij}=\frac{4(t'_{ij})^2}{U},
\end{align}
with diagonal hoppings $t'_{ij}=t_\pm=t'\pm\delta$ giving $J_\pm=4(t'\pm\delta)^2/U$.
We assume $(\pi,\pi)$ N\'eel order and rotate spins on sublattice $B$ by $\pi$ (about $x$), so that for $i\in A$, $j\in B$,
\begin{align}
\bS_i\cdot\bS_j
\rightarrow
- S_i^z S_j^z+\frac12(S_i^+S_j^+ + S_i^-S_j^-),
\end{align}
whereas for same-sublattice bonds (AA or BB) $\bS_i\cdot\bS_j=S_i^zS_j^z+\frac12(S_i^+S_j^-+S_i^-S_j^+)$ remains unchanged.
Introducing Holstein–Primakoff bosons,
\begin{align}
i\in A:\;\;
&S_i^z=S-a_i^\dagger a_i,\quad
S_i^+=\sqrt{2S}\,a_i,\quad
S_i^-=\sqrt{2S}\,a_i^\dagger,
\\
j\in B:\;\;
&S_j^z=S-b_j^\dagger b_j,\quad
S_j^+=\sqrt{2S}\,b_j,\quad
S_j^-=\sqrt{2S}\,b_j^\dagger,
\end{align}
and retaining terms up to quadratic order, a NN $A$--$B$ bond gives
\begin{align}
H^{(1)}_{ij}
&=J_1\Big[-S_i^zS_j^z+\frac12(S_i^+S_j^+ + S_i^-S_j^-)\Big]\nonumber\\
&\approx -J_1S^2
+J_1S\big(a_i^\dagger a_i+b_j^\dagger b_j\big)
+J_1S\big(a_i b_j+a_i^\dagger b_j^\dagger\big),
\label{eq:NNbond}
\end{align}
while a diagonal bond on $A$--$A$ (or $B$--$B$) gives
\begin{align}
H^{(2)}_{ij}
&=J_{ij}\Big[S_i^zS_j^z+\frac12(S_i^+S_j^-+S_i^-S_j^+)\Big]\nonumber\\
&\approx J_{ij}S^2
-J_{ij}S\big(\alpha_i^\dagger \alpha_i+\alpha_j^\dagger \alpha_j\big)
+J_{ij}S\big(\alpha_i^\dagger \alpha_j+\alpha_j^\dagger \alpha_i\big),
\label{eq:diagbond}
\end{align}
where $\alpha=a$ if $i,j\in A$ and $\alpha=b$ if $i,j\in B$. We can now use the Fourier Transformation as 
\begin{align}
a_i=\frac{1}{\sqrt{N}}\sum_{\bk}a_{\bk}e^{i\bk\cdot\br_i},\qquad
b_j=\frac{1}{\sqrt{N}}\sum_{\bk}b_{\bk}e^{i\bk\cdot\br_j},\end{align}

and similarly for $a_i^\dagger,b_j^\dagger$.

\subsubsection{NN contribution.}
From \eqref{eq:NNbond}, summing the number terms over all NN bonds,
\begin{align}
\sum_{\langle ij\rangle} a_i^\dagger a_i
=\sum_{i\in A}\Big(\sum_{j\in{\rm NN}(i)}1\Big)a_i^\dagger a_i
=z_1\sum_{i\in A}a_i^\dagger a_i,
\qquad z_1=4,
\end{align}
and analogously $\sum_{\langle ij\rangle} b_j^\dagger b_j=4\sum_{j\in B}b_j^\dagger b_j$. Hence
\begin{align}
J_1S\sum_{\langle ij\rangle}(a_i^\dagger a_i+b_j^\dagger b_j)
=4SJ_1\sum_{i\in A}a_i^\dagger a_i+4SJ_1\sum_{j\in B}b_j^\dagger b_j
=4SJ_1\sum_{\bk}(a_{\bk}^\dagger a_{\bk}+b_{\bk}^\dagger b_{\bk}).
\end{align}

We can write NN sites as $j=i+\delta$ with $\delta\in\{(\pm1,0),(0,\pm1)\}$:
\begin{align}
\sum_{\langle ij\rangle}a_i b_j
=\sum_{i\in A}\sum_{\delta\in{\rm NN}} a_i\,b_{i+\delta}.
\end{align}
Using the Fourier transforms:
\begin{align}
a_i\,b_{i+\delta}
=\frac{1}{N}\sum_{\bk,\bq}a_{\bk}b_{\bq}\,
e^{i(\bk+\bq)\cdot\br_i}\,e^{i\bq\cdot\delta},
\end{align}
and using $\sum_{i\in A}e^{i(\bk+\bq)\cdot\br_i}=N\delta_{\bq,-\bk}$ to obtain
\begin{align}
\sum_{\langle ij\rangle}a_i b_j
=\sum_{\bk}a_{\bk}b_{-\bk}\sum_{\delta\in{\rm NN}}e^{i\bk\cdot\delta}
=\sum_{\bk}a_{\bk}b_{-\bk}\,2(\cos k_x+\cos k_y).
\end{align}
Therefore the NN pairing term becomes
\begin{align}
J_1S\sum_{\langle ij\rangle}(a_i b_j+a_i^\dagger b_j^\dagger)
=\sum_{\bk}B(\bk)\big(a_{\bk}b_{-\bk}+a_{\bk}^\dagger b_{-\bk}^\dagger\big),
\qquad
B(\bk)=2SJ_1(\cos k_x+\cos k_y).
\end{align}

\subsubsection{NNN contribution.}
From \eqref{eq:diagbond}, the diagonal onsite contribution on sublattice $A$ is
\begin{align}
-\,S\sum_{\langle\!\langle ij\rangle\!\rangle\subset A}J_{ij}(a_i^\dagger a_i+a_j^\dagger a_j)
=-S\sum_{i\in A}a_i^\dagger a_i\Big(\sum_{j\in{\rm diag}(i)}J_{ij}\Big).
\end{align}
For the checkerboard plaquette modulation we take: on $A$ sites the two $\pm(1,1)$ diagonals carry $J_+$ and the two $\pm(1,-1)$ diagonals carry $J_-$, so $\sum_{j\in{\rm diag}(i)}J_{ij}=2J_+ +2J_-$. Hence
\begin{align}
-\,S\sum_{\langle\!\langle ij\rangle\!\rangle\subset A}J_{ij}(a_i^\dagger a_i+a_j^\dagger a_j)
=-2S(J_+ +J_-)\sum_{i\in A}a_i^\dagger a_i
=-2S(J_+ +J_-)\sum_{\bk}a_{\bk}^\dagger a_{\bk}.
\end{align}
The same diagonal onsite shift holds on sublattice $B$:
$-2S(J_+ +J_-)\sum_{\bk}b_{\bk}^\dagger b_{\bk}$.
For Diagonal hopping terms, we can
split the diagonal vectors into $D_+=\{\pm(1,1)\}$ and $D_-=\{\pm(1,-1)\}$. For $A$--$A$ bonds,
\begin{align}
S\sum_{\langle\!\langle ij\rangle\!\rangle\subset A}J_{ij}(a_i^\dagger a_j+a_j^\dagger a_i)
&=S\sum_{i\in A}\Big[
J_+\sum_{\delta\in D_+}(a_i^\dagger a_{i+\delta}+a_{i+\delta}^\dagger a_i)
+J_-\sum_{\delta\in D_-}(a_i^\dagger a_{i+\delta}+a_{i+\delta}^\dagger a_i)
\Big].
\end{align}
Using
\begin{align}
\sum_{i\in A}(a_i^\dagger a_{i+\delta}+a_{i+\delta}^\dagger a_i)
=2\sum_{\bk}a_{\bk}^\dagger a_{\bk}\cos(\bk\cdot\delta),
\end{align}
and $\sum_{\delta\in D_+}\cos(\bk\cdot\delta)=2\cos(k_x+k_y)$,
$\sum_{\delta\in D_-}\cos(\bk\cdot\delta)=2\cos(k_x-k_y)$, we obtain
\begin{align}
S\sum_{\langle\!\langle ij\rangle\!\rangle\subset A}J_{ij}(a_i^\dagger a_j+a_j^\dagger a_i)
=2S\sum_{\bk}a_{\bk}^\dagger a_{\bk}\Big[J_+\cos(k_x+k_y)+J_-\cos(k_x-k_y)\Big].
\end{align}
On sublattice $B$ the couplings are interchanged ($J_+\leftrightarrow J_-$), giving
\begin{align}
S\sum_{\langle\!\langle ij\rangle\!\rangle\subset B}J_{ij}(b_i^\dagger b_j+b_j^\dagger b_i)
=2S\sum_{\bk}b_{\bk}^\dagger b_{\bk}\Big[J_-\cos(k_x+k_y)+J_+\cos(k_x-k_y)\Big].
\end{align}

Henceforth,
 Collecting NN and diagonal contributions,
\begin{align}
H_2
=E_0+\sum_{\bk}\Big[
A_a(\bk)\,a_{\bk}^\dagger a_{\bk}
+A_b(\bk)\,b_{\bk}^\dagger b_{\bk}
+B(\bk)\,(a_{\bk}b_{-\bk}+a_{\bk}^\dagger b_{-\bk}^\dagger)
\Big],
\end{align}
with
\begin{align}
A_a(\bk)&=4SJ_1-2S(J_+ + J_-)+2S\Big[J_+\cos(k_x+k_y)+J_-\cos(k_x-k_y)\Big],\\
A_b(\bk)&=4SJ_1-2S(J_+ + J_-)+2S\Big[J_-\cos(k_x+k_y)+J_+\cos(k_x-k_y)\Big],\\
B(\bk)&=2SJ_1(\cos k_x+\cos k_y).
\end{align}

\subsection{Magnon Spectrum}

We need to diagonalize the Hamiltonian in the Nambu spinor basis. Define $\Phi_{\bk}=(a_{\bk},\,b_{-\bk}^\dagger)^T$, $\eta={\rm diag}(1,-1)$ and
$\mathcal H_{\bk}=\begin{psmallmatrix}A_a(\bk)&B(\bk)\\ B(\bk)&A_b(\bk)\end{psmallmatrix}$, so that
$H_2=E_0+\tfrac12\sum_{\bk}\Phi_{\bk}^\dagger\mathcal H_{\bk}\Phi_{\bk}$ up to a constant.
The magnon energies follow from $\eta\mathcal H_{\bk}u=\omega u$, i.e.
\begin{align}
\det(\eta\mathcal H_{\bk}-\omega I)
=\omega^2-(A_a-A_b)\omega-(A_aA_b-B^2)=0.
\end{align}
Writing $D(\bk)=\sqrt{(A_a(\bk)+A_b(\bk))^2-4B(\bk)^2}$, the two positive branches are
\begin{align}
\omega_{+}(\bk)=\frac{D(\bk)+(A_a(\bk)-A_b(\bk))}{2},\qquad
\omega_{-}(\bk)=\frac{D(\bk)-(A_a(\bk)-A_b(\bk))}{2},
\end{align}
while the full BdG eigenvalue set is $\{\pm\omega_+(\bk),\pm\omega_-(\bk)\}$.

\section{Random phase approximation}
\label{app:RPA}

We can compare the results of our SBMFT calculations with less sophisticated methods like the random phase approximation (RPA). We thereby calculate the elementary excitations within the AM state via the bare dynamic particle-hole susceptibility
\begin{equation}
    \chi^0_{\{o_i,s_i\}}(\boldsymbol{q}, \omega) = 
    - \int_\mathrm{BZ}\,\frac{\mathrm{d}\boldsymbol k}{V_\mathrm{BZ}} \sum_{nm} \frac{f(\beta E_n(\boldsymbol k + \boldsymbol q)) - f(\beta E_m(\boldsymbol k))}{E_n(\boldsymbol k + \boldsymbol q) - E_m(\boldsymbol k) + \omega - \text{i} \eta}
    \,
    [u^n_{o_1, s_1}(\boldsymbol k + \boldsymbol q) \,
    u^m_{o_2, s_2}(\boldsymbol k)]^* \,
    u^n_{o_3, s_3}(\boldsymbol k + \boldsymbol q) \,
    u^m_{o_4, s_4}(\boldsymbol k) \ ,
\end{equation}
where $\beta$ is the inverse temperature, $f(\beta E)$ is the Fermi distribution and the sum over $n,m$ runs over all bands in the system.
The electronic eigenstates satisfying $H u^n_{o,s}(\boldsymbol{k}) = E_n(\boldsymbol{k}) \, u^n_{o,s}(\boldsymbol{k})$ with eigenvalues defined by \cref{eq:minimal_bands_compact} give the orbital to band transformation of the Lindhard function.
The momentum space integral is evaluated on a mesh in the BZ of volume $V_\mathrm{BZ}$ on a $1000 \times 1000$ grid employing the implementation described in Ref.~\cite{Duerrnagel2022}.
To ensure numerical convergence, we have introduced a small infrared regulator $\eta = 10^{-2}$ and used $T = 0.005$ in accordance with the SBMFT calculations. 

We then calculate corrections to this non-interacting susceptibility via the Bethe-Salpeter equation effectively corresponding to summing up all particle-hole ladder diagrams leading to the interacting RPA spin susceptibility $\chi^\text{RPA}(\boldsymbol{Q}, \omega)$.
Taking into account only the onsite repulsion $U$, the different spin sectors decouple~\cite{Costa2025}, such that we can simply write
\begin{equation}
    \chi^\pm_\text{RPA}(\boldsymbol{q}, \omega) =
    (1 - U \chi^\pm_0(\boldsymbol{q}, \omega))^{-1} \chi^\pm_0(\boldsymbol{q}, \omega) \ ,
\end{equation}
where the matrix valued quantities are obtained by considering the spin correlation functions on the two sublattice sites
\begin{equation}
    [\chi^\pm_0]_{o_1 o_2}(\boldsymbol{q}, \omega) =
    \langle T S^\pm_{o_1}(- \boldsymbol{q}) S^\mp_{o_2}(\boldsymbol{q}) \rangle = 
        \sum_{\{s_i\}} \delta_{o_2 o_3} \delta_{o_1 o_4} \,
        \sigma^\pm_{s_1 s_4} \sigma^\mp_{s_2 s_3} \,
        \chi^0_{\{o_i, s_i\}} (\boldsymbol{q}, \omega)\ .
\end{equation}
Here, we use $\sigma^\pm = \sigma^x \pm \text{i} \sigma^y$.
This is equivalent to considering Gaussian fluctuations around the static mean-field solution. Crucially, this procedure neglects fluctuation feedback on the MF state as well as inter-channel feedback between spin and charge fluctuations.

\begin{figure}[t]
    \centering
    \includegraphics[width=.5\linewidth]{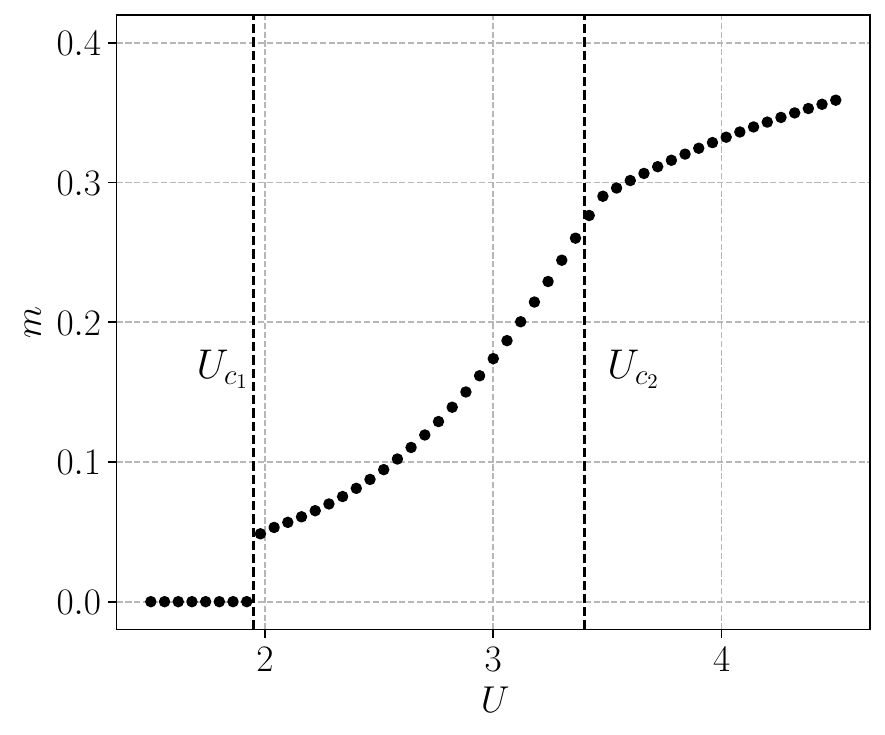}
    \caption{Magnetization curve as obtained by solving the self-consistent gap equation \cref{eq:gap_equation} with the critical interaction values for the AM transition $U_{c_1}$ and MIT $U_{c_2}$ indicated.}
    \label{fig:mf_results}
\end{figure}

To obtain a consistent description within the RPA, we can hence not employ the ordered state from SBMFT for the subsequent RPA analysis but have to obtain the static MF state self-consistently. To this end, we fix the MF order parameter to a staggered sublattice magnetization density
\begin{equation}
    M_{o_1 o_2} = M \tau^z_{o_1o_2} =
    \frac{m U}{N} \sum_{\boldsymbol{k} s_1 s_2} \langle c_{o_1s_1}^\dagger \sigma^z_{s_1s_2} \tau^z_{o_1o_2} 
    c_{o_2s_2}^\nodag \rangle
\end{equation}
with $\sigma^z$ ($\tau^z$) given by the third Pauli matrix in spin (sublattice) space respectively, for which we obtain the self-consistency equation
\begin{equation}
    M = - \frac{U}{2 N} \sum_{\boldsymbol k n} f(\beta E_n(\boldsymbol k)) \,\frac{\partial E_n(\boldsymbol k)}{\partial M} \,.
    \label{eq:gap_equation}
\end{equation}
at fixed filling $n = 1$ with $E_n(\boldsymbol{k})$ being the eigenenergies of the MF Hamiltonian in \cref{eqn:H_MF}.
The overall magnetization strength $m$ is then obtained via $m = M / U$ and displayed in \cref{fig:mf_results}.

\begin{figure}[t]
    \centering
    \includegraphics[width=.5\linewidth]{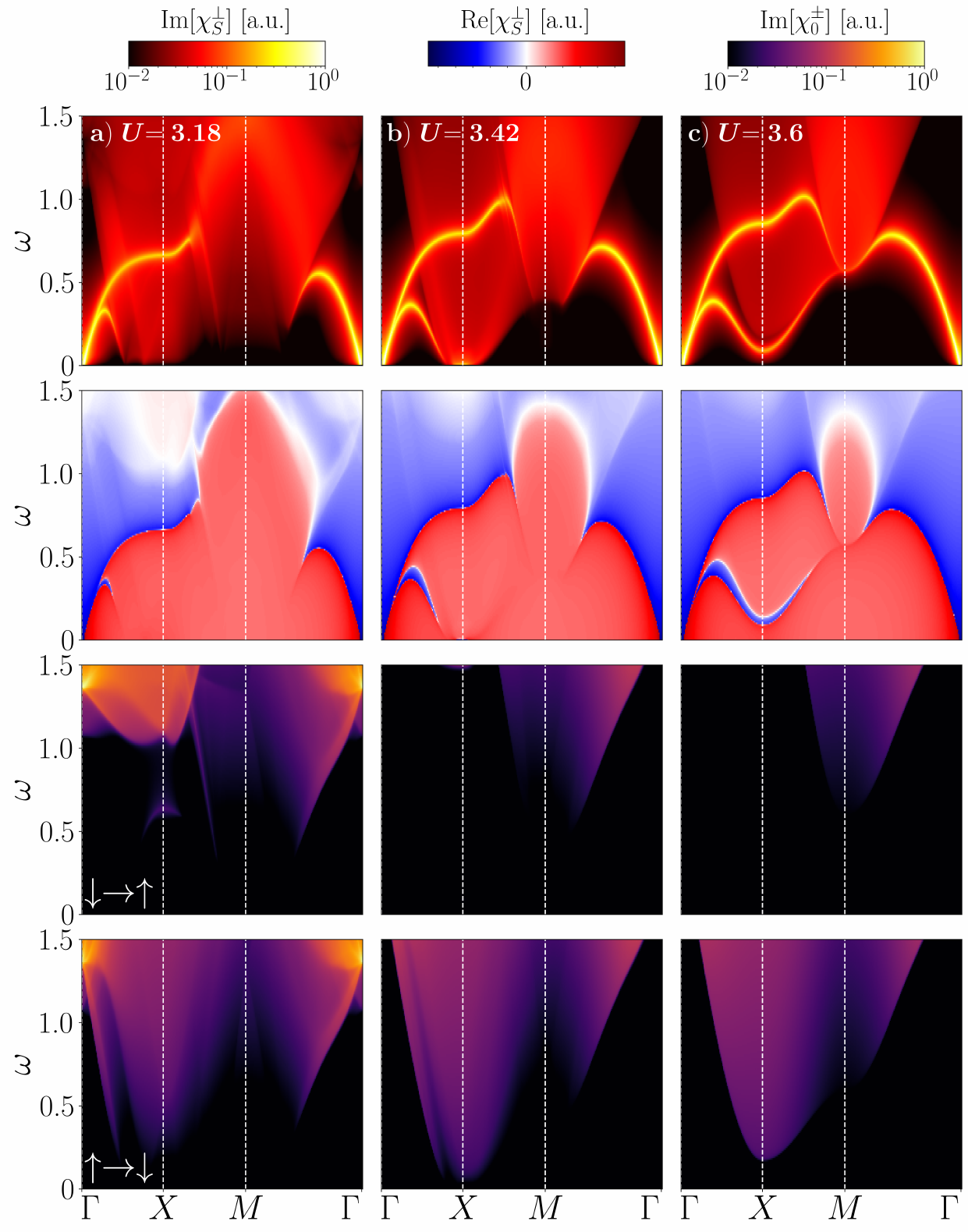}
    
    \caption{Magnon spectra from RPA across the MIT.
    The presentation of the data is the same as in \cref{fig:afm-am} in the main text, but at fixed $\delta = 0.2 t$ at various interactions.
    The different values correspond in the self-consistent MF to phase space points a) deep inside the AM--M phase, b) directly at the MIT, and  c) close to the MIT in the AM--I phase.}
    \label{fig:magnons_rpa}
\end{figure}

The resulting magnon dispersions in the different AM regimes are depicted in \cref{fig:magnons_rpa} and reveal a very good qualitative agreement with the magnon spectra in \cref{fig:sus_chirality}. In particular, the sharp but heavily deformed magnon mode at the $X$ point is recovered right until the MIT, while the degenerate magnons at $M$ suffer from considerable damping already in the insulating case.
This is somehow expected: Since the model Hamiltonian \cref{eq:hamiltonian_tt2d} provides the simplest model to host AM phenomenology, the quasi-particle band structure from self-consistent MF and SBMFT are very similar, since self-energy effects are not capable of substantially altering the structure of the electronic Green's functions due to the high symmetry of the model. Hence, the Stoner continua in both cases are very similar and consequently the dressed magnon modes obtained thereof.

However, the self-consistent MF is heavily underestimating the involved critical interactions $U_{c_1}$ and $U_{c_2}$ and at the same time predicts much larger magnetization values compared to the SBMFT calculations (cf.~\cref{fig:maxwell_combined}). This is a well known feature of MF theories since quantum fluctuation effects are not treated sufficiently in this approach. In particular, the lack of other competing fluctuation channels renders the intriguing phase separation regime at the MIT inaccessible.
This highlights the insufficiency of MF methods and derived methods like DFT+U to quantitatively assess the critical scales and magnetization~\cite{Smolyanyuk2024} and accordingly the chiral magnon splitting~\cite{Costa2025} in actual materials. 

\end{document}